\documentstyle[preprint,aps]{revtex}

\begin{document}

\title{Resonant tunneling through a small quantum dot coupled to
superconducting leads}

\author{A. Levy Yeyati, J.C. Cuevas, 
A. L\'{o}pez-D\'avalos\cite{Arturo} and A. Mart\'{\i}n-Rodero}

\address{
Departamento de F\'\i sica  Te\'orica de la Materia Condensada C-V.\\
Universidad Aut\'onoma de Madrid,
E-28049 Madrid, Spain.}

\maketitle

\begin{abstract}
We address the problem of non-linear transport through discrete 
electronic levels in a small quantum dot 
coupled to superconducting electrodes. In our approach 
the low temperature $I-V$ characteristics can be
calculated including all multiple 
quasi-particle and Andreev processes. 
The limit of very weak coupling
to the leads and large charging energies is briefly analyzed 
comparing the calculated lineshapes of the $I-V$ curves 
with recent experimental results.
When the coupling to the
leads increases and Coulomb blockade effects can be neglected, the 
combination of multiple Andreev processes and
resonant transmission gives rise to a rich subgap structure
which largely differs from the one found in the more studied S-N-S systems. 
We concentrate on this regime showing
how multiple processes can be included within a
simple sequential tunneling picture qualitatively explaining the 
subgap structure.  
We suggest an experimental set-up where the predicted effects could be
observed.
\end{abstract}


Resonant tunneling through small systems characterized by 
discrete electronic states weakly coupled to
metallic leads has been the object of extensive studies for nearly a
decade \cite{Kastner}. This physical situation is, in a sense, common to 
a large
variety of systems ranging from semiconductor quantum dots
\cite{artificial}, small metallic islands
\cite{island} or atomic impurities \cite{impurity}, which can exhibit 
characteristic phenomena like
Coulomb-blockade and the Kondo effect.
More recently there has been a growing interest in the
special case in which these systems include superconducting parts
bringing the possibility of observing new
effects associated with the superconducting state.
While a large part of these efforts have been devoted to understand the
interplay between charging and paring effects in ``large'' mesoscopic
superconducting islands with a mean level spacing 
much smaller than the energy gap \cite{parity}, 
less attention has been paid to the case where discrete levels can be
resolved \cite{Khlus}. In this communication we
shall consider this latter situation for the case where both leads
are superconducting. 

A system of this kind has been recently investigated experimentally
by Ralph et al. who could resolve individual electronic states in the
tunneling through a nanometer $Al$ particle weakly coupled to
superconducting leads \cite{Ralph}.  
For these small particle sizes and
extremely weak coupling, both the charging energy, $E_C$, and the
single-particle level spacing, $\delta$, 
are much larger than the gap parameter of the superconducting leads,
$\Delta$. 
In this regime the
contribution of Andreev processes can be shown to be negligible and
only single quasi-particle processes need to be considered. On the other
hand, for a system with a larger coupling to the leads and smaller charging
energies, multiple Andreev reflections (MAR) may become very important for
determining the transport properties. One would expect that 
the combination of MAR with
resonant transmission through the discrete levels would
give rise to a rich subgap structure (SGS) in the $I-V$ characteristics,
containing new features when compared to the more studied case of
S-N-S junctions.
An ideal experimental set-up for exploring these effects would be one in
which both the resonant level position and the coupling to the
leads could be modified in a controlled way like in a normal
artificial atom \cite{artificial}. This situation is 
represented schematically in Fig. 1. 
 
The aim of this communication is to give a theoretical analysis 
of the non-linear dc characteristics for systems of the kind discussed
above. The analysis is non-perturbative in the coupling to the leads and
therefore valid both for the weak and strong coupling regimes.
In the weak coupling case our theory naturally explains
the observed lineshapes 
\cite{Ralph} without the need to resort to 
a phenomenological broadening of the BCS density of states.
We shall mainly concentrate in
the case where the resonant level lies within the
energy gap region and the Coulomb blockade can be neglected, for which
a novel subgap structure is predicted to appear. 

As mentioned above, a small quantum dot connected to superconducting
leads is characterized by the interplay of many different energy
scales. In order to analyze in detail the basic phenomena that can be
observed in the case of large mean level spacing ($\delta \gg \Delta$),
we shall restrict the present theoretical
discussion to the simplest case where transport takes place through a
single resonant level. 
For describing this physical situation we use the following model
Hamiltonian

\begin{equation}
\hat{H}= \hat{H}_L + \hat{H}_R + \sum_{\nu,\sigma} t_{\nu} 
(\hat{c}^{\dagger}_{\nu \sigma} \hat{c}_{0 \sigma} +
\hat{c}^{\dagger}_{0 \sigma} \hat{c}_{\nu \sigma}) +
\sum_{\sigma} \epsilon_0 \hat{n}_{0 \sigma} +
U \hat{n}_{0 \uparrow} \hat{n}_{0 \downarrow}
\end{equation}

\noindent
where $\hat{H}_L$ and $\hat{H}_R$ are BCS Hamiltonians
describing the left and right superconducting leads,
characterized by gap parameters $\Delta_L= \Delta_R= \Delta$; 
$\epsilon_0$ is the bare resonant level position,
$t_{\nu}$ with $\nu = L, R$ are 
hopping parameters which connect the level to the left and
right leads, and the $U$ term describes the
intralevel Coulomb repulsion. 
This parameter is basically the dot 
charging energy, $E_C$, and is related to the total dot capacitance 
$C$, by $U \sim e^2/2C$ \cite{capacitance}.
For the case of superconducting leads 
it is convenient to choose a
gauge in which an applied bias voltage is introduced through time
dependent phase factors modulating the hopping parameters \cite{tocho} as 
$t_{\nu} \rightarrow t_{\nu} \exp(i e V_{\nu} \tau /\hbar)$, where 
$V_{\nu}$ are the voltage drops between the leads and the
central region. 
Hamiltonian (1) is nothing but an Anderson model which 
has been extensively
used for studying the case of a small dot coupled to normal leads
\cite{dot}. For the
subsequent discussion it is convenient to introduce the normal elastic
tunneling rates $\Gamma_{\nu} = \pi  |t_{\nu}|^2 \rho_{\nu}(\mu)$,
where $\rho_{\nu}(\mu)$ are the normal spectral densities of the
leads at the Fermi level.

There are two different regimes in which correlation effects associated
with the $U$-term in Hamiltonian (1) can either be neglected or taken
into account in a simple way, and yet lead to a
non-trivial behavior. The first regime corresponds to a case where the
couplings to the leads are not extremely small and the dot capacitance is
large enough as to smear out the Coulomb
blockade effect ($\Gamma_{L,R} \sim U$). In this case 
the system behaves as if there were a single spin-degenerate
resonant level at $\epsilon \simeq \epsilon_0 + U <n_0>$, 
i.e. a restricted Hartree approximation on the $U$-term
would be reasonable. When this effective level lies within the
superconducting gap both single quasiparticle and Andreev
processes give an important contribution to the subgap $I-V$ structure. 
The second regime would correspond to the
experimental conditions of Ref. \cite{Ralph} in which $U  \gg \Delta
\gg \Gamma_{L,R}$. Double occupancy of the resonant level becomes then
very unlikely, Andreev processes are strongly suppressed and thus only
single-quasiparticle processes have to be considered. This situation can
be simply simulated by replacing the isolated dot Hamiltonian by a single
non-degenerate effective level. 
The analysis of other regimes,
where correlations effects could play a relevant role (like for 
instance in the Kondo effect) would be given in a forthcoming
publication.

The transport properties of this model can be obtained using the same
approach as in Refs. \cite{tocho} which is based on
non-equilibrium Green functions techniques. As discussed in these references,
this method is based on an expansion of the relevant Green functions 
in terms of the hopping elements coupling the dot to the leads.
When both electrodes are superconducting, due to the presence of
MAR, this expansion leads to an infinite set
of algebraic equations for the Green functions, which can be solved
using recursive techniques as discussed in detail in Refs. \cite{tocho}. 
As a consequence of this infinite series of MAR,  
the average current, $I(\tau)$, contains 
all harmonics of the Josephson frequency $\omega_0 = 2eV/ \hbar$, i.e. 
the current can be written as 
$I(\tau)= \sum_n I_n \exp(in \omega_0 \tau)$. We shall
concentrate here in its dc part $I_0$. 
Further details of this formalism, as applied to the present case, will
be given elsewhere.

Let us first analyze the simplest case $U \gg \Delta \gg \Gamma$. As
mentioned above the theory is then greatly simplifyed by suppression of
Andreev reflections. The dc current is then given by the resonant
tunneling like expression

\begin{equation}
I_0(V)= \frac{4e}{h} \int^{\infty}_{-\infty}
d \omega \, \frac{\Gamma^S_{L}(\omega) \; 
\Gamma^S_{R}(\omega)}
{(\omega - \epsilon)^2 + \left[ \Gamma^S_{L}(\omega) +
\Gamma^S_{R}(\omega) \right]^2} \;
\left[ n_F(\omega - eV/2) - n_F(\omega + eV/2) \right] ,
\end{equation}

\noindent
where the superconducting tunneling rates $\Gamma^S_{L,R}$ are defined
as $\Gamma^S_{L,R}(\omega)= \Gamma_{L,R} \tilde{\rho}^S(\omega \pm eV/2)$, 
$\tilde{\rho}^S$ being the 
corresponding dimensionless BCS spectral density given by 
$\tilde{\rho}^S = |\omega| /\sqrt{\omega^2 - \Delta^2}$, and 
$n_F(\omega)$ is the 
Fermi distribution function. In this expression $\epsilon$  is the
effective resonant level position in which the charging effects have
been included.

In Fig. 2 $I_0(V)$ is shown for decreasing coupling 
to the leads. Notice that the lineshape progressively resembles a BCS
spectral density in agreement with the experimental observations
\cite{Ralph}. In Ref. \cite{Ralph} special attention was given 
to the behavior
of the $I-V$ characteristics near the threshold voltage, in
particular, to the broadening and reduced amplitude with respect to
a simple BCS spectral density. In fact, our theory predicts a finite
height and width of these resonances without having to resort to the
introduction of any phenomenological broadening parameter.
This is a simple consequence of having a small but finite coupling to
the leads. 
Eq. (2) in the limit of vanishing coupling reduces to 
$ I_0(V) \sim \Gamma^S_{L}(\epsilon) \Gamma^S_{R}(\epsilon)
/( \Gamma^S_{L}(\epsilon) 
+ \Gamma^S_{R}(\epsilon)) $,
which coincides with the result one would expect from a sequential 
tunneling picture \cite{sequential}. 
The coherent processes taken into account in Eq. (2) are 
responsible for the rounding off of the resonant peaks, as illustrated in 
Fig. 2. We have estimated that the experimental situation in Ref. 
\cite{Ralph} would approximately correspond to 
the case plotted as a full line in Fig. 2.

Let us next consider the case where $U \sim \Gamma_{L,R}$, which, as
discussed above, can be
described by a single spin-degenerate effective level. 
In contrast to the previous case, now the effect of MAR become crucial
and the full formalism is needed for the calculation of the current.
The importance of MAR in this case is
clearly displayed in the subgap structure exhibited in 
the $I-V$ curves shown in Fig. 3. This figure
corresponds to a symmetric 
case where the level position $\epsilon$, is fixed
at $\epsilon=0$, while $\Gamma = \Gamma_L = \Gamma_R$ varies between
$\Gamma \gg \Delta$ and $\Gamma < \Delta$. As can be observed, in the limit
$\Gamma \gg \Delta$ SGS is absent, the relevant feature being a
saturation of the current at $I_0 \sim 4e \Delta/h$ for $V \rightarrow 0$.
This feature is commonly referred to as a ``foot" in the context of
S-N-S structures \cite{Gunsen}. In this regime we recover the result for 
a ballistic
single-mode superconducting quantum point contact as obtained recently 
by different authors
\cite{tocho,Averin}. The departure from the S-N-S behavior becomes 
more apparent for decreasing $\Gamma$, where a
progressively pronounced SGS appears. Notice that, in contrast to what 
is found
in S-N-S structures with a continuum of states in the central region, 
in the present case
the $I_0(V)$ curve {\it itself} exhibits oscillations with the
concomitant appearance of negative differential conductance.

The particular SGS found in this system is related to the influence of
resonant tunneling on the MAR processes.
In fact, 
in the limit of $\Gamma \ll \Delta$ the position and shape of the subgap
current peaks can be understood by means of the following simple
picture. When $eV < 2 \Delta$ current between the superconducting leads
can flow due to MAR processes. As in the case of S-S or S-N-S junctions, there
appear jumps in the subgap $I-V$ characteristics at voltages
corresponding to the opening of a new Andreev channel.
However, the amplitude of these subgap processes
is greatly modified by the presence of a resonant level between the
leads, in such a way that only those MAR ``trajectories" that connect
the resonant level to the leads spectral densities give a significant
contribution to the current. 
The inset in Fig. 4a illustrates the 2nd order Andreev trajectory which 
gives the
dominant contribution to the current when $2 \Delta/3 < eV < 2 \Delta$.
As $eV$ decreases towards $2 \Delta/3$, the energies of the initial and
final states on this trajectory approach the gap edges,
which results in a BCS DOS-like shape of the current peak at
$eV = 2 \Delta/3$ (see Fig. 4a).

This simple picture enables us to evaluate the SGS at $\Gamma \ll
\Delta$ analytically.
The first step is to identify the {\it generalized} tunneling rates 
$\Gamma^{(in)}_n (\omega)$ and $\Gamma^{(out)}_n (\omega)$ 
associated with the $n$ order Andreev process connecting the dot to the
leads. Here the $\Gamma^{(in)}_n$'s give the probability of an electron 
or a hole to get into the dot as an electron, while the $\Gamma^{(out)}_n$'s
correspond to the complementary processes where one electron leaves the
dot and reaches the leads as an electron or as a hole. A simple analysis
yields, for the symmetric case $\Gamma = \Gamma_L = \Gamma_R$,
the following expressions

\begin{equation}
\Gamma^{(out,in)}_n(\omega) = \Gamma^{2n+1} \tilde{\rho}^S(\omega \pm
(2n+1) eV/2) 
\prod^n_{j=1} \left| \frac{f(\omega \pm (2j-1) eV/2)}{ \omega - 
(-1)^j \epsilon
\pm jeV} \right|^2 
\end{equation}

\noindent
where $f(\omega)$ is the dimensionless BCS pairing amplitude of the 
uncoupled leads defined as $f(\omega)= \Delta / \pi \sqrt{\Delta^2 -
\omega^2}$. Notice that $\Gamma^2 |f(\omega)|^2$ is 
the Andreev reflection probability at the lowest order in $\Gamma$,
while the denominator in Eq. (3) is related to the transmission
probability through the dot.
The total current can then be computed as the sum of the
contributions due to all possible combinations of {\it in} and
{\it out} processes. Every contribution must be weighted by
the total charge which is transferred in the combined process. 
For instance, the process depicted as an inset in Fig. 4a is a
combination of two processes of the first order and has an associated
charge of $3e$. The 
resulting expression for the current is

\begin{equation}
I_0(V) = \frac{8 e \pi}{h} \sum_{n,m} (n + m + 1)
\frac{\Gamma^{(in)}_n (\epsilon)
\Gamma^{(out)}_m (\epsilon)}{\Gamma^{(in)}_n (\epsilon) + 
\Gamma^{(out)}_m (\epsilon)}.  
\end{equation} 

Fig. 4 illustrates in further detail the SGS of the 
$I-V$ curves for $\Gamma \ll \Delta$.
For comparison the results given by Eq. (4) 
are also shown. As can be observed, this simple approximation
fairly reproduces the exact numerical results in this small $\Gamma$
limit.  
The rounding off of the peaks which is
observed for increasing $\Gamma$ (see Fig. 3) is due to off resonant
processes as discussed for the case of large charging energy. In the 
limit $\Gamma \sim \Delta$ the sequential tunneling picture breaks down
due to the interference among the different multiple processes.

For the general case with $\epsilon \ne 0$, the SGS becomes more complex
due to the appearance of additional resonances. This is illustrated in
Fig. 4b for the case $\epsilon = 0.2 \Delta$ and $\Gamma = 5 \times
10^{-2} \Delta$. As can be deduced from the sequential formulae of Eq. (4), 
resonances appear both at 
$eV^+_n = 2 (\Delta + \epsilon)/(2n + 1)$ and 
$eV^-_n = 2 (\Delta - \epsilon)/(2n + 1)$ corresponding to processes in
which the inital or final states are at the gap edges. The sequential 
picture also predicts the appearance of resonances at $j eV = 2 \epsilon$
due to resonant coupling between electron and hole states. In the exact
numerical results the resonances are somewhat shifted with respect to
these predictions and some of
them are difficult to resolve. Nevertheless, it should be stressed that
the main qualitative features of the exact SGS are already contained
in Eq. (4).                

As has been already mentioned, an ideal experimental set-up to study the
interplay between resonant tunneling and MAR would be an 
``artificial-atom" with superconducting leads as represented in Fig. 1.
In this type of structure, for a quantum dot area of $\sim (100 nm)^2$,
the mean level spacing would be around $2.7 meV$ \cite{capacitance}, which 
is  much larger than the superconducting gap on the leads if these   
were made of $Al$ ($\Delta_{Al} \sim
0.18 meV$). On the other hand, the coupling to the leads could be given
any desired value 
by varying the conductance of the two point contacts, and the charging
energy could also in principle be changed by varying the different
capacitances between the dot and the surrounding metallic leads and gates. 
Thus, the conditions for observing the 
$\Gamma \sim U < \Delta$ could be attainable.
In fact, a system of these characteristics would not be very different
from the superconducting quantum point contact developed by Takayanagi et al.
\cite{Takayanagi}.

In conclusion, we have presented model calculations for the transport
through discrete resonant levels coupled to superconducting
leads. We have briefly analyzed the case corresponding
to a large charging energy and a extremely weak coupling, finding good
agreement with the experimental results of Ref. \cite{Ralph}.
The main part of this work has been concentrated in the study of the
case $U \sim \Gamma < \Delta$, where Coulomb blockade effects can be
neglected. For this regime we have shown how the interplay
between resonant tunneling and multiple Andreev reflection processes 
give rise to a novel
subgap structure in the $I-V$ curves, which could be in 
principle detected if the adequate conditions are met in the experiments. 
We have shown how the effects of MAR can be taken into account in a 
sequential tunneling picture by the introduction of generalized
tunneling rates. This opens the
possibility of analyzing more complex situations in which the mean
level spacing is comparable to the superconducting gap and multiple 
resonances are involved. Work along these lines is under progress.

Support by Spanish CICYT (Contract No. PB93-0260). One of us (A.L.D.)
acknowledges partial financial support by ``Instituto Nicol\'as
Cabrera".

\begin{figure}
\caption{Schematical representation of a quantum dot defined in a 
2D electron gas and coupled to superconducting leads} 
\end{figure}

\begin{figure}
\caption{Zero temperature $I-V$ characteristics corresponding to the case
of large charging energy and extremely small coupling. 
Full line for $\Gamma_L = 5 \times 10^{-3}
\Delta$, dashed line for $\Gamma_L = 10^{-3} \Delta$ and dotted line for
$\Gamma_L = 2 \times 10^{-4} \Delta$. In all cases $\Gamma_R = 4 \Gamma_L$ 
and $\epsilon = 5 \Delta$.} 
\end{figure}

\begin{figure}
\caption{Zero temperature $I-V$ characteristics corresponding to the case
where Coulomb blockade is absent for different values of $\Gamma =
\Gamma_L = \Gamma_R$. The effective resonant level is $\epsilon = 0$.}
\end{figure}

\begin{figure}
\caption{Detail of the subgap structure
for $\Gamma = 5
\times 10^{-2} \Delta$ with $\epsilon = 0$ (a) and $\epsilon = 0.2
\Delta$ (b). 
Full line: complete numerical calculation; dotted line: sequential 
approximation discussed in the text. 
The inset represents a typical resonant ``trajectory'' mediated by 
two Andreev reflections. The arrows indicate the position of the first
resonances for $\epsilon = 0.2 \Delta$.} 
\end{figure}

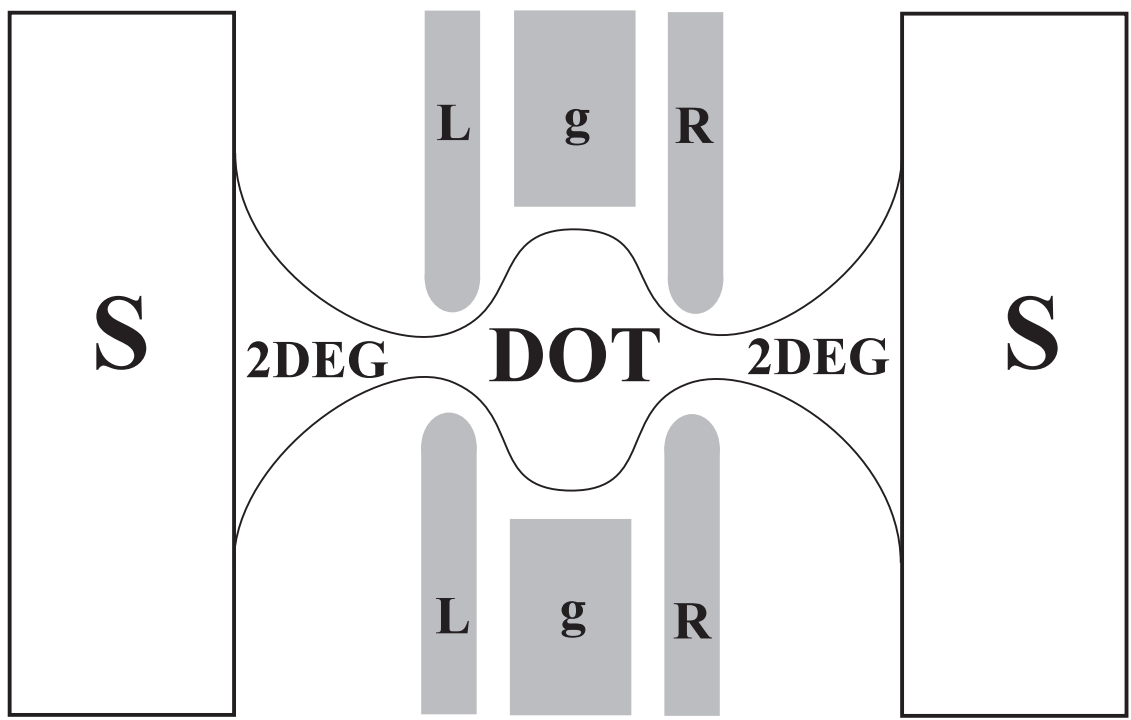
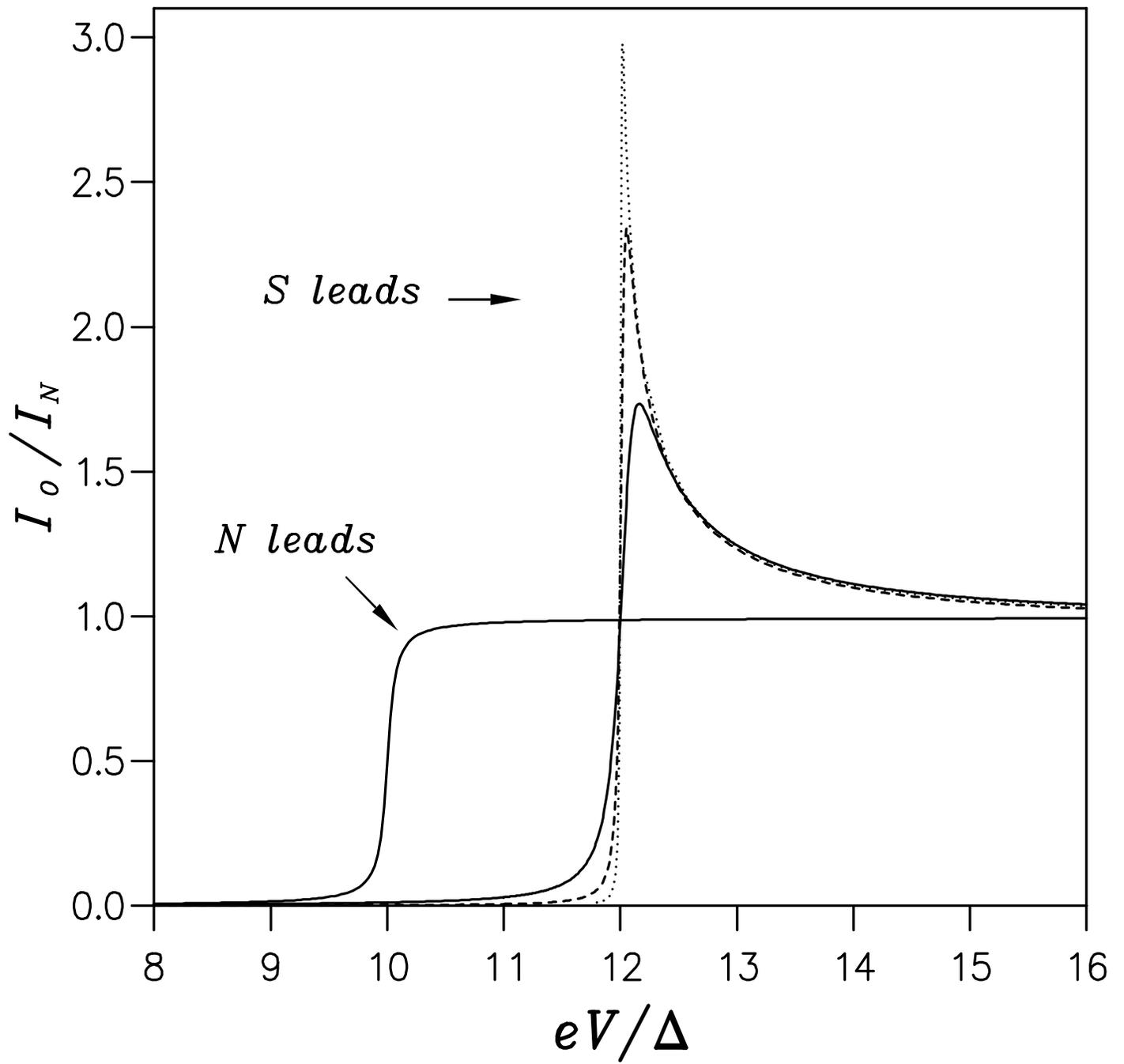
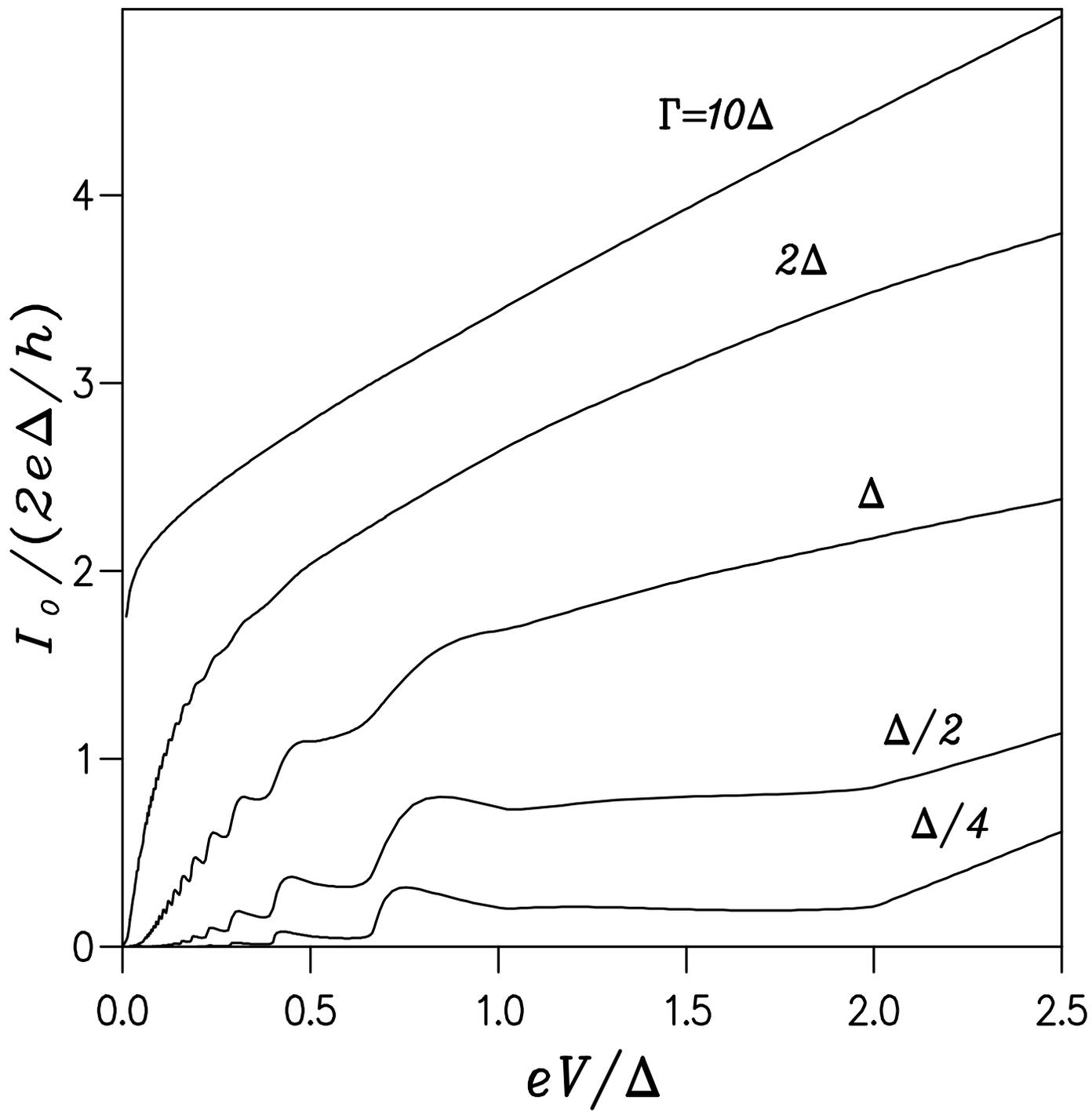
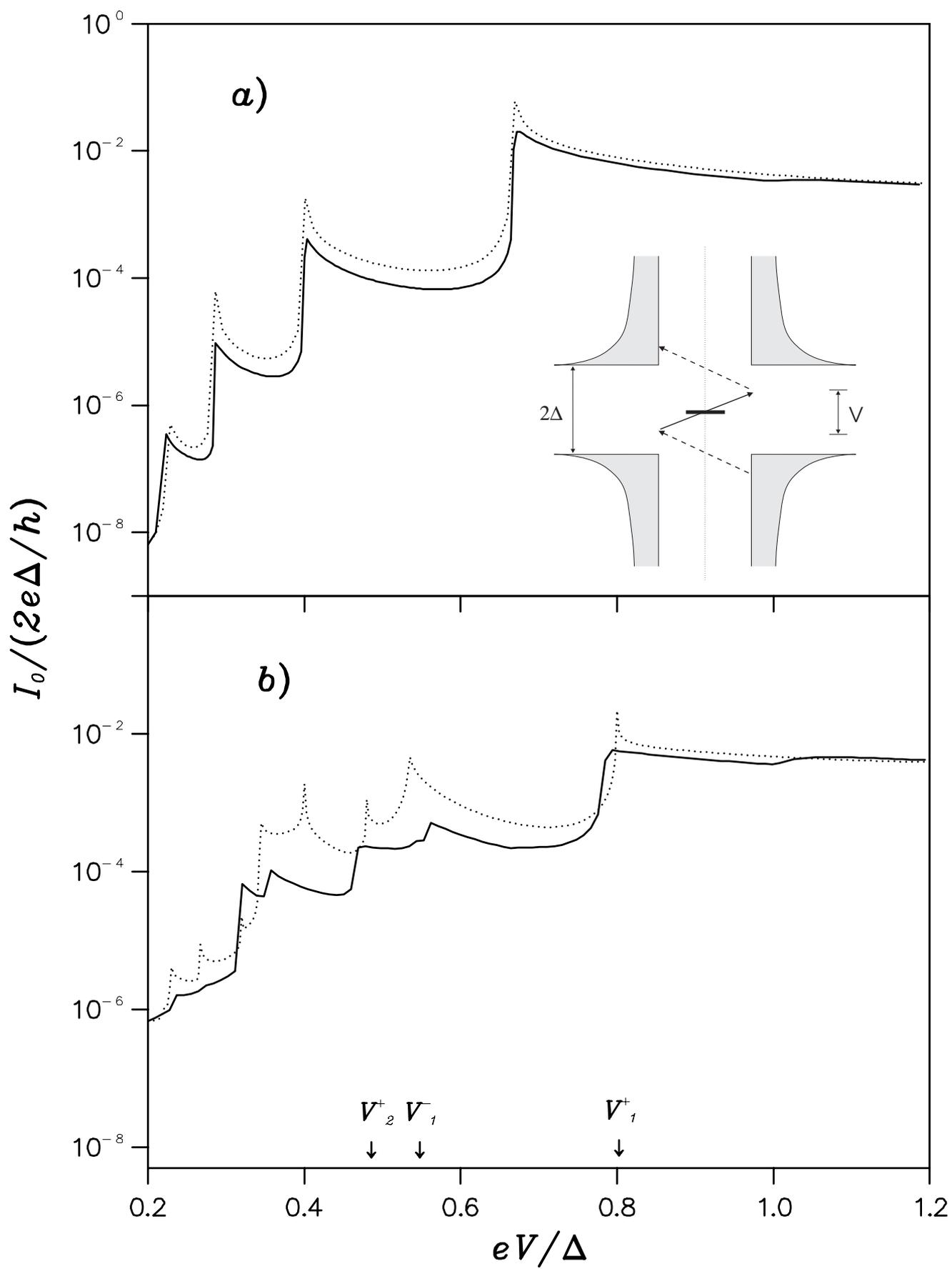
\end{document}